\documentclass[11pt,twoside]{article}
\usepackage[pdftex]{graphicx}
\usepackage{amsmath}
\usepackage{amssymb}
\usepackage{cite}
\begin{document}
\newcommand{\pst}{\hspace*{1.5em}}

\newcommand{\rigmark}{\em Journal of Russian Laser Research}
\newcommand{\lemark}{\em Volume 35, Number 1, 2014}

\newcommand{\be}{\begin{equation}}
\newcommand{\ee}{\end{equation}}
\newcommand{\bm}{\boldmath}
\newcommand{\ds}{\displaystyle}
\newcommand{\bea}{\begin{eqnarray}}
\newcommand{\eea}{\end{eqnarray}}
\newcommand{\ba}{\begin{array}}
\newcommand{\ea}{\end{array}}
\newcommand{\arcsinh}{\mathop{\rm arcsinh}\nolimits}
\newcommand{\arctanh}{\mathop{\rm arctanh}\nolimits}
\newcommand{\bc}{\begin{center}}
\newcommand{\ec}{\end{center}}

\thispagestyle{plain}

\label{sh}


\begin{center} {\Large \bf
\begin{tabular}{c}
Tomographic and improved subadditivity conditions\\
for two qubits and qudit with $j=3/2$
\end{tabular}
}
\end{center}

\bigskip

\bigskip

\begin{center} {\bf
V. N. Chernega and O. V. Man'ko$^*$ }
\end{center}

\medskip

\begin{center}
{\it P.~N.~Lebedev Physical Institute, Russian Academy of Sciences\\
Leninskii Prospect 53, Moscow 119991, Russia}
\end{center}

\smallskip

\smallskip

$^*$Corresponding author e-mail:~~~omanko@sci.lebedev.ru

\begin{abstract}\noindent
New entropic inequality for quantum and tomographic Shannon information for system of two qubits is obtained. The inequality relating quantum information and spin-tomographic information for particle with spin $j=3/2$ is found. The method to extend the obtained new entropic and information inequalities for one qudit and arbitrary composite system of qudits is suggested.
\end{abstract}

\medskip


\section{Introduction}
\pst The states of classical systems with fluctuating observables due to interaction with environment are described by probability distributions. The probability distributions associated with one random variable $q$ in the case of finite number of outcomes equal $N$ are identified with probability vectors
${\bf p}=(p_1,p_2,\ldots,p_N)$ where $1\geq p_k\geq 0$, $k=1,2,\ldots,N$ and $\sum_{k=1}^{N} p_k=1$. The states of quantum systems, e.g. qudit or spin $j$ states where $j=0,1/2,1,\ldots$ are described by density $N\times N$-matrix \cite{Landau,Landau1,vonNeuman,vonNeuman1} $\rho_{mm'}=\langle m|\hat\rho|m'\rangle$ where spin projections $m,m'$ take the values $-j,-j+1,\ldots,j-1,j$. The diagonal elements of the density matrix $p_m=\rho_{mm}$ can be considered as components of the probability vector ${\bf p}=(p_{-j},p_{-j+1},\ldots,p_{j-1},p_{j})$, where $1\geq p_m\geq 0$, $\sum_{m=-j}^{j} p_m=1$. For several $M=2,3, \ldots$ random classical variables the probability distributions are joint probability distributions ${\cal P}(n_1,n_2,\ldots,n_M)\geq 0$ such that
\begin{equation}\label{eq.1}
\sum_{n_1=1}^{N_1}\sum_{n_2=1}^{N_2}\ldots\sum_{n_M=1}^{N_M} {\cal P}(n_1,n_2,\ldots,n_M)=1.
\end{equation}
For several $M$ quantum particles, e.g. for composite system of qudits or composite system of $M$ particles with spins $j_1,j_2,\ldots,j_M$ the state density operator $\hat\rho$ has the density matrix $\rho_{{\bf m},{\bf m'}}=\langle{\bf m}|\hat\rho|{\bf m'}\rangle$ where ${\bf m}=(m_1,m_2,\ldots,m_M),$ ${\bf m'}=(m_1',m_2',\ldots,m_M'),$ are the vectors with components corresponding to spin $j_k$ projections $m_k=-j_k,-j_k+1,\ldots,j_k$ and $k=1,2,\ldots,M$. The diagonal elements of the density matrix $\rho_{{\bf mm}}=\langle{\bf m}|\hat\rho|{\bf m}\rangle$ give the joint probability distribution ${\cal P}(m_1,m_2,\ldots,m_M)\geq 0$ satisfying the normalization condition
\begin{equation}\label{eq.2}
\sum_{m_1=-j_1}^{j_1}\sum_{m_2=-j_2}^{j_2}\ldots\sum_{m_M=-j_M}^{j_M}{\cal P}(m_1,m_2,\ldots,m_M)=1.
\end{equation}
The statistical properties of quantum observables were shown to be described in terms of standard probability theory formalism in \cite{CherManko6}.

Each probability distribution is characterized by the Shannon entropy \cite{Shanon}
\begin{equation}\label{eq.3}
H_{{\bf p}}=-\sum_{k=1}^{N}p_k\ln p_k\geq 0
\end{equation}
for one probability vector and
\begin{equation}\label{eq.4}
H_{{\cal P}}=-\sum_{n_1=1}^{N_1}\sum_{n_2=1}^{N_2}\ldots \sum_{n_M=1}^{N_M}{\cal P}(n_1,n_2,\ldots,n_M)\ln
{\cal P}(n_1,n_2,\ldots,n_M)\geq 0
\end{equation}
for the joint probability distribution ${\cal P}(n_1,n_2,\ldots,n_M)$. For quantum states the entropy is defined in terms of density operator $\hat\rho$ as von Neumann entropy
\begin{equation}\label{eq.5}
S=-\mbox{Tr}\hat\rho\ln\hat\rho.
\end{equation}
If the density matrix of quantum state $\rho_{{\bf mm'}}$ which has the properties to be Hermitian, i.e. $(\rho_{{\bf mm'}})^+=(\rho_{{\bf mm'}})$ and trace-class, i.e. $\mbox{Tr}\rho=1$ as well to have only nonnegative eigenvalues, i.e. $\hat\rho\geq 0$, is diagonalized, the entropy (\ref{eq.5}) can be expressed in the form of Shannon entropy determined by the eigenvalues of the density operators identified with the joint probability distribution ${\cal P}(n_1,n_2,\ldots,n_M)$. The entropies determined by classical probability vectors and joint probability distributions obey some inequalities. The entropies determined by the density operators also obey other inequalities \cite{LiebSeiringer,LeibJMathPhys,Petz,ECarlen,Wehrl,VBRUskai,Holevo,FoundPhysRita,MendesJRLR,PhysScrRita,30DecLosAlamPrepr}. The inequalities called "subadditivity condition", "stronger subadditivity condition" and strong subadditivity condition are considered to be valid for composite classical and quantum systems. Recently, new representation of quantum mechanics called "tomographic probability representation of quantum mechanics" was introduced \cite{Mancini96} (see also reviews \cite{OlgaJRLR1997,Ibort,MankoMarmo}). In this representation the density operators are mapped onto standard probability distributions called "quantum state tomograms". The connection between tomographic schemes and star--product quantization procedure was investigated in \cite{OlgaMankoMarmoPhSc,OlgaBeppeJPA,MarmoKapuschik,OlgaPatr}. The general geometrical relations of quantum tomographic probability distributions with all the other possible quasidistributions were found in \cite{Mendes}. The review of tomographic representation of quantum and classical mechanics can be found in \cite{Ibort,MankoMarmo,OlgaVova}. There exist important relations of entropies associated with the tomographic probability distributions and von Neumann entropies \cite{MendesJRLR,ZborovskiPilMano}.  These relations were used to introduce the notion of quantum tomographic discord \cite{RitaJRLR,Yurkevich}. On the other hand there exist mathematical inequalities for entropies associated with the nonnegative Hermitian matrices \cite{Kim,LiebSeiringer,LeibJMathPhys,Petz,ECarlen,Wehrl,VBRUskai,Holevo,FoundPhysRita,MendesJRLR,PhysScrRita,30DecLosAlamPrepr}. 

The aim of this article is to find the connection between the mathematical inequalities and new tomographic inequalities for entropies of physical system states. Also we extend the inequalities considered usually for bipartite system to the case of systems without subsystems. To do this we apply the qubit (qudit) portrait method \cite{Vovf,Zupo,30DecLosAlamPrepr} in the spirit of the application of the method to get subadditivity and strong subadditivity condition to qudit states \cite{Vova}. 

The paper is organised as follows. In Sec.2 we demonstrate our method and obtain the result on example of two qubit states. The qudit states for the case $j=3/2$ are studied in Sec.3. In Sec.4 we describe application of the method to get new inequalities for generic qudit states. In Sec. 5 we study some nonlinear maps of density matrices and get improved subadditivity condition for $j=3/2$. In Sec.6 we present conclusions and perspectives.

\section{Two-qubit entropic inequalities}
In this section we consider in detail the examples of system state in four-dimensional Hilbert space. The density matrix of a system state in a four-dimensional Hilbert-space reads
\begin{equation}\label{eq.1a}
\rho=
\left(
  \begin{array}{cccc}
    \rho_{11} & \rho_{12} &\rho_{13}  &\rho_{14}  \\
    \rho_{21} & \rho_{22} &\rho_{23} &\rho_{24} \\
    \rho_{31} & \rho_{32} & \rho_{33} & \rho_{34} \\
    \rho_{41} & \rho_{42} &\rho_{43}  &\rho_{44}  \\
  \end{array}
\right), \quad \rho_{j k}=\langle j|\hat\rho|k\rangle, \quad j,k=1,2,3,4.
\end{equation}
The matrix has the properties $\rho_{j k}=\rho^*_{kj}$, $\mbox{Tr}\rho=1$. The eigenvalues of the density operator $\hat\rho$ and its $4\times4$ matrix $\rho$ are nonnegative numbers $(\rho_1,\rho_2,\rho_3,\rho_4)={\bf p}$. These properties do not define what physical system has the state with the density matrix (\ref{eq.1a}). To clarify this issue one must define the properties of basis $|k\rangle$ ($k=1,2,3,4$) in the Hilbert space. For example if this matrix corresponds to two-qubit states it can be rewritten in the form
\begin{equation}\label{eq.2a}
 \rho=
\left(
  \begin{array}{cccc}
    \rho_{1/2\,1/2,1/2\,1/2} & \rho_{1/2\,1/2,1/2\,-1/2} &\rho_{1/2\,1/2,-1/2\,1/2}  &\rho_{1/2\,1/2,-1/2\,-1/2}  \\
    \rho_{1/2\,-1/2,1/2\,1/2} & \rho_{1/2\,-1/2,1/2\,-1/2} &\rho_{1/2\,-1/2,-1/2\,1/2} &\rho_{1/2\,-1/2,-1/2\,-1/2} \\
    \rho_{-1/2\,1/2,1/2\,1/2} & \rho_{-1/2\,1/2,1/2\,-1/2} & \rho_{-1/2\,1/2,-1/2\,1/2} & \rho_{-1/2\,1/2,-1/2\,-1/2} \\
    \rho_{-1/2\,-1/2,1/2\,1/2} & \rho_{-1/2\,-1/2,1/2\,-1/2} &\rho_{-1/2\,-1/2,-1/2\,1/2} &\rho_{-1/2\,-1/2,-1/2\,-1/2} \\
  \end{array}
\right).
\end{equation}
We used the notation for the basis vector $|k\rangle$ in the form $|m_1m_2\rangle$, the vector is eigenvector of the two operators $\hat J_{z1}=\hat J_z\otimes\hat1_2$ and $\hat J_{z2}=\hat 1_2\otimes\hat J_z$ where the operator $\hat J_z$ has the $2\times2$-matrix $J_z=\frac{1}{2}\sigma_z$, the matrix $\sigma_z$ is the Pauli matrix, i.e.,
$\sigma_z=\left(
            \begin{array}{cc}
              1 & 0 \\
              0 & -1 \\
            \end{array}\right).$ Thus the $4\times4$-matrices of operators $\hat J_z\otimes\hat 1_2$ and $\hat 1_2\otimes\hat J_z$ read
\begin{equation}\label{eq.3a}
  J_{z1}=J_z\otimes1_2=\left(
                         \begin{array}{cccc}
                           1 & 0 & 0 & 0 \\
                           0 & 1 & 0 & 0 \\
                           0 & 0 & -1 & 0 \\
                           0 & 0 & 0 & -1 \\
                         \end{array}
                       \right), \quad J_{z2}=1_2\otimes J_z=\left(
                         \begin{array}{cccc}
                           1 & 0 & 0 & 0 \\
                           0 & -1 & 0 & 0 \\
                           0 & 0 & 1 & 0 \\
                           0 & 0 & 0 & -1 \\
                         \end{array}
                       \right).
\end{equation}
These matrices commute. It means that both observables $\hat J_{z1}$ and $\hat J_{z2}$ can be measured simultaneously. They have common set of eigenvectors $|m_1m_2\rangle$. Thus one has $\hat J_{z1}|m_1m_2\rangle=m_1|m_1m_2\rangle,$ $\hat J_{z2}|m_1m_2\rangle=m_2|m_1m_2\rangle$, where $m_1,m_2$ take the values $\pm 1/2$. In fact, the possibility to consider the matrix  $\rho$ (\ref{eq.1a}) as the density matrix of two-qubit state is based on possibility to make the following invertable map of natural numbers onto pairs of fractions 
\begin{equation}\label{eq.4a}
  1\Longleftrightarrow1/2\,1/2,\quad 2\Longleftrightarrow1/2\,-1/2,\quad 3\Longleftrightarrow-1/2\,1/2,\quad 4\Longleftrightarrow-1/2\,-1/2.
\end{equation}
The map (\ref{eq.4a}) means that the first natural numbers $1,2,3,4$ can be coded by all the pairs of possible spin projections $m,m'$ on arbitrary quantization direction, e.g., $z$-axes , in the system of two qubits. Since the matrices (\ref{eq.1a}) and (\ref{eq.2a}) are the same matrices, all their corresponding matrix elements are equal, e.g.
\begin{equation}\label{eq.5a}
\rho_{11}=\rho_{1/2\,1/2,1/2\,1/2},\quad\rho_{12}=\rho_{1/2\,1/2,1/2\,-1/2},\ldots,  \rho_{44}=\rho_{-1/2\,-1/2,-1/2\,-1/2}. 
\end{equation}
Representation of the density matrix (\ref{eq.1a}) in the form (\ref{eq.2a}) provides possibility to apply obvious for two-qubit state procedure of taking the partial trace of the density matrix. In fact for two-qubit state the density matrix can be denoted as the matrix $\rho\equiv\rho(1,2)$ of the operator $\hat\rho(1,2)$ of bipartite system with matrix elements $\rho_{m_1\,m_2,m_1'\,m_2'}$ given by (\ref{eq.2a}). The partial trace procedure means the following positive map of the matrix (\ref{eq.2a}) (and also matrix (\ref{eq.1a}))
\begin{equation}\label{eq.6a}
\rho(1,2)\rightarrow\rho(1)=\mbox{Tr}_2\rho(1,2),\quad  \rho(1,2)\rightarrow\rho(2)=\mbox{Tr}_1\rho(1,2). 
\end{equation}
Explicit form of taking partial trace is the map
\begin{equation}\label{eq.7a}
\sum_{m_2=-1/2}^{1/2}\rho(1,2)_{m_1m_2m_1'm_2}=\rho(1)_{m_1m_1'}, 
\end{equation}
and another map
\begin{equation}\label{eq.8a}
\sum_{m_1=-1/2}^{1/2}\rho(1,2)_{m_1m_2m_1m_2'}=\rho(2)_{m_2m_2'}.
\end{equation}
We get using (\ref{eq.7a}), (\ref{eq.8a})
\begin{equation}\label{eq.9a}
\rho(1)=\left(
          \begin{array}{cc}
            \rho(1)_{1/2\,1/2} &\rho(1)_{1/2\,-1/2} \\
            \rho(1)_{-1/2\,1/2} &\rho(1)_{-1/2\,-1/2} \\
          \end{array}
        \right), \quad\rho(2)=\left(
          \begin{array}{cc}
            \rho(2)_{1/2\,1/2} &\rho(2)_{1/2\,-1/2} \\
            \rho(2)_{-1/2\,1/2} &\rho(2)_{-1/2\,-1/2} \\
          \end{array}
        \right).
\end{equation}
The obtained matrix elements read
\begin{eqnarray}
&&\rho(1)_{1/2\,1/2}=\rho_{1/2\,1/2,1/2\,1/2}+\rho_{1/2\,-1/2,1/2\,-1/2},\nonumber\\&&\rho(1)_{1/2\,-1/2}= \rho(1)^*_{-1/2\,1/2}=
\rho_{1/2\,1/2,-1/2\,1/2}+\rho_{1/2\,-1/2,-1/2\,-1/2},\nonumber\\
&&\rho(1)_{-1/2\,-1/2}=\rho_{-1/2\,1/2,-1/2\,1/2}+\rho_{-1/2\,-1/2,-1/2\,-1/2}\label{eq.10a}
\end{eqnarray}
and
\begin{eqnarray}
&&\rho(2)_{1/2\,1/2}=\rho_{1/2\,1/2,1/2\,1/2}+\rho_{-1/2\,1/2,-1/2\,1/2},\nonumber\\ &&\rho(2)_{1/2\,-1/2}= \rho(2)^*_{-1/2\,1/2}=
\rho_{1/2\,1/2,1/2\,-1/2}+\rho_{-1/2\,1/2,-1/2\,-1/2},\nonumber\\
&&\rho(2)_{-1/2\,-1/2}=\rho_{1/2\,-1/2,1/2\,-1/2}+\rho_{-1/2\,-1/2,-1/2\,-1/2}.\label{eq.11a}
\end{eqnarray}
One can interpreter the maps in the form of the positive maps of $4\times4$ matrices
\begin{equation}\label{eq.12a}
\rho(1,2)\rightarrow\left(
                      \begin{array}{cc}
                        \rho(1) & 0 \\
                        0 & 0 \\
                      \end{array}
                    \right), \quad \rho(1,2)\rightarrow\left(
                      \begin{array}{cc}
                        \rho(2) & 0 \\
                        0 & 0 \\
                      \end{array}
                    \right).
\end{equation}
On the other hand the maps obtained by the partial trace procedure can be applied directly to the matrix (\ref{eq.1a}).
In this case the maps read
\begin{equation}\label{eq.13a}
\rho\rightarrow\left(
                 \begin{array}{cc}
                   \rho_1 & 0 \\
                   0 & 0 \\
                 \end{array}
               \right); \quad \rho\rightarrow\left(
                                \begin{array}{cc}
                                  \rho_2 & 0 \\
                                  0 & 0 \\
                                \end{array}
                              \right),
\end{equation}
where
\begin{equation}\label{eq.14a}
\rho_1=\left(
             \begin{array}{cc}
               \rho_{11}+\rho_{22} &\rho_{13}+\rho_{24} \\
               \rho_{31}+\rho_{42} &\rho_{33}+\rho_{44} \\
             \end{array}
           \right),\quad 
\rho_2=\left(
             \begin{array}{cc}
               \rho_{11}+\rho_{33} &\rho_{12}+\rho_{34} \\
               \rho_{21}+\rho_{43} &\rho_{22}+\rho_{44} \\
             \end{array}
           \right).
\end{equation}
It is worthy to note that all 24 permutations of the four numbers $1,2,3,4\rightarrow1_p,2_p,3_p,4_p$ which provide permutations of basis vectors $|j\rangle\rightarrow|j_p\rangle$ yield the positive maps of the initial matrix $\rho\rightarrow\rho_p$. The matrix elements of the matrix $(\rho_p)_{jk}$ become $\rho_{j_pk_p}$. Then the maps (\ref{eq.13a}) and (\ref{eq.14a}) become the positive maps of initial matrix $\rho$ onto the matrices $\rho_1^{(p)}$, $\rho_2^{(p)}$ where these matrices have the matrix elements with index permutations $j\rightarrow j_p$, $k\rightarrow k_p$. Due to invertable coding of the numbers $1,2,3,4$ by pairs of the spin projections $m,m'$ the obvious positive maps of two-qubit matrix given by (\ref{eq.9a}) provide after applying the number permutation tool  new maps of the two-qubit density matrix which are not obtained by simple partial tracing.

Standard two-qubit entropic inequalities which are nonnegativity of von Neumann entropies of any qubit system and subadditivity condition in the system of two qubits read
\begin{equation}\label{eq.15a}
-\mbox{Tr}_1\rho(1)\ln\rho(1)\geq0;\quad-\mbox{Tr}_2\rho(2)\ln\rho(2)\geq0;\quad
-\mbox{Tr}_{12}\rho(1,2)\ln\rho(1,2)\geq0
\end{equation}
and
\begin{equation}\label{eq.16a}
-\mbox{Tr}_1\rho(1)\ln\rho(1)-\mbox{Tr}_2\rho(2)\ln\rho(2)\geq-\mbox{Tr}_{12}\rho(1,2)\ln\rho(1,2).
\end{equation}
Here the density matrices of first qubit $\rho(1)$ and second qubit $\rho(2)$ given by (\ref{eq.9a}) with matrix elements (\ref{eq.10a}), (\ref{eq.11a}) are obtained by procedure of partial tracing of the density matrix $\rho(1,2)$ given by (\ref{eq.2a}) using the map of indices (\ref{eq.5a}). On the other hand the density matrix $\rho$ given by (\ref{eq.2a}) is identical to the matrix (\ref{eq.12a}). In view of this the inequalities (\ref{eq.15a}), (\ref{eq.16a}) can be written in the form
\begin{equation}\label{eq.17a}
-\mbox{Tr}\left(
            \begin{array}{cc}
              \rho_{11}+\rho_{22} &\rho_{13}+\rho_{24} \\
              \rho_{42}+\rho_{31} &\rho_{33}+\rho_{44} \\
            \end{array}
          \right)\ln\left(
            \begin{array}{cc}
              \rho_{11}+\rho_{22} &\rho_{13}+\rho_{24} \\
              \rho_{42}+\rho_{31} &\rho_{33}+\rho_{44} \\
            \end{array}
          \right)\geq0.
\end{equation}
Analogous nonnegativity condition for von Neumann entropy associated with the matrix $\rho_2$ given by (\ref{eq.14a}) is also  valid. The analog of subadditivity condition (\ref{eq.16a}) can be written for the matrix $\rho$ given by (\ref{eq.1a}). It looks as follows
\begin{eqnarray}
&&-\mbox{Tr}\left(
  \begin{array}{cccc}
    \rho_{11} & \rho_{12} &\rho_{13}  &\rho_{14}  \\
    \rho_{21} & \rho_{22} &\rho_{23} &\rho_{24} \\
    \rho_{31} & \rho_{32} & \rho_{33} & \rho_{34} \\
    \rho_{41} & \rho_{42} &\rho_{43}  &\rho_{44}  \\
  \end{array}
\right)\ln\left(
  \begin{array}{cccc}
    \rho_{11} & \rho_{12} &\rho_{13}  &\rho_{14}  \\
    \rho_{21} & \rho_{22} &\rho_{23} &\rho_{24} \\
    \rho_{31} & \rho_{32} & \rho_{33} & \rho_{34} \\
    \rho_{41} & \rho_{42} &\rho_{43}  &\rho_{44}  \\
  \end{array}
\right)\leq\nonumber\\
&&-\mbox{Tr}\left(
            \begin{array}{cc}
              \rho_{11}+\rho_{22} &\rho_{13}+\rho_{24} \\
              \rho_{42}+\rho_{31} &\rho_{33}+\rho_{44} \\
            \end{array}
          \right)\ln\left(
            \begin{array}{cc}
              \rho_{11}+\rho_{22} &\rho_{13}+\rho_{24} \\
              \rho_{42}+\rho_{31} &\rho_{33}+\rho_{44} \\
            \end{array}
          \right)\nonumber\\
&&-\mbox{Tr}\left(
            \begin{array}{cc}
              \rho_{11}+\rho_{33} &\rho_{12}+\rho_{24} \\
              \rho_{21}+\rho_{43} &\rho_{22}+\rho_{44} \\
            \end{array}
          \right)\ln\left(
            \begin{array}{cc}
              \rho_{11}+\rho_{33} &\rho_{12}+\rho_{34} \\
              \rho_{21}+\rho_{43} &\rho_{22}+\rho_{44} \\
            \end{array}
          \right).\label{eq.18a}
\end{eqnarray}
There are 24 inequalities which can be obtained from (\ref{eq.18a}) for the matrix elements of the matrix $\rho$ (\ref{eq.1a}) by all the permutations of the numbers $1,2,3,4.$ Thus we got entropic inequalities for arbitrary Hermitian nonnegative $4\times4$-matrix $\rho$ with $\mbox{Tr}\rho=1$. We obtained these inequalities using identity of the mathematical structure of this matrix with the density matrix of two-qubit system.

\section{The entropic inequalities for the density matrix of qudit with $j=3/2$}
Let us consider the density matrix of the state of qudit corresponding to spin $j=3/2$. This Hermitian nonnegative $4\times4$-matrix $\rho_{mm'}^{3/2}$ has the form
\begin{equation}\label{eq.19a}
\rho^{3/2}=\left(\begin{array}{cccc}
\rho_{3/2\,\,3/2} & \rho_{3/2\,\,1/2} &\rho_{3/2\,\,-1/2}  &\rho_{3/2\,\,-3/2}  \\
\rho_{1/2\,\,3/2} & \rho_{1/2\,\,1/2} &\rho_{1/2\,\,-1/2} &\rho_{1/2\,\,3/2} \\
\rho_{-1/2\,\,3/2} & \rho_{-1/2\,\,1/2} & \rho_{-1/2\,\,-1/2} & \rho_{-1/2\,\,-3/2} \\
\rho_{-3/2\,\,3/2} & \rho_{-3/2\,\,1/2} &\rho_{-3/2\,\,-1/2}  &\rho_{-3/2\,\,-3/2}  \\
\end{array}\right).
\end{equation}
One has property $\mbox{Tr}\rho^{3/2}=1.$ The indices $m,m'$ in the matrix $\rho_{mm'}^{3/2}=\langle m|\hat\rho^{3/2}|m'\rangle$ are spin projections taking values $3/2,1/2,-1/2,-3/2$. Let us now introduce the invertable map of four numbers $3/2\leftrightarrow1,\,1/2\leftrightarrow2,\,-1/2\leftrightarrow3,\,-3/2\leftrightarrow4.$
Applying this map to the matrix (\ref{eq.19a}) we obtain the form of the matrix $\rho^{3/2}$ identical to the matrix (\ref{eq.1a}). It means that all the equalities and inequalities known for the matrix (\ref{eq.1a}) are valid for the matrix (\ref{eq.19a}). Thus we can obtain new inequalities for the density matrix $\rho^{3/2}$ of the qudit state given by Eq. (\ref{eq.19a}) by making the indices map, e.g. in (\ref{eq.17a}) and (\ref{eq.18a}).

We will write these new inequalities explicitly. We have inequality $-\mbox{Tr}\rho^{3/2}\ln\rho^{3/2}\geq0$. In explicit form one has also the inequality 
\begin{equation}\label{eq.20a}
-\mbox{Tr}\left(\begin{array}{cc}
\rho_{3/2\,\,3/2}+ \rho_{1/2\,\,1/2} &\rho_{3/2\,\,-1/2}+ \rho_{1/2\,\,-3/2} \\
\rho_{-1/2\,\,3/2}+\rho_{-3/2\,\,1/2} &\rho_{-1/2\,\,-1/2}+ \rho_{-3/2\,\,-3/2}\\ \end{array}\right) 
\ln \left(\begin{array}{cc}
\rho_{3/2\,\,3/2}+ \rho_{1/2\,\,1/2} &\rho_{3/2\,\,-1/2}+ \rho_{1/2\,\,-3/2} \\
\rho_{-1/2\,\,3/2}+\rho_{-3/2\,\,1/2} &\rho_{-1/2\,\,-1/2}+ \rho_{-3/2\,\,-3/2}\\
\end{array}\right)\geq 0,
\end{equation}            
analogous to nonnegativity of the von Neumann entropy of the "qubit-subsystem" state, but the qudit with $j=3/2$ does not have such subsystem. Analogous entropic nonnegativity condition can be written by using the introduced map of indices in the matrix $\rho_2$ given by (\ref{eq.14a}). The new "subadditivity condition" for qubit states with $j=3/2$ has the explicit form
\begin{eqnarray}
&&-\mbox{Tr}\left(\begin{array}{cccc}
\rho_{3/2\,\,3/2} & \rho_{3/2\,\,1/2} &\rho_{3/2\,\,-1/2}  &\rho_{3/2\,\,-3/2}  \\
\rho_{1/2\,\,3/2} & \rho_{1/2\,\,1/2} &\rho_{1/2\,\,-1/2} &\rho_{1/2\,\,3/2} \\
\rho_{-1/2\,\,3/2} & \rho_{-1/2\,\,1/2} & \rho_{-1/2\,\,-1/2} & \rho_{-1/2\,\,-3/2} \\
\rho_{-3/2\,\,3/2} & \rho_{-3/2\,\,1/2} &\rho_{-3/2\,\,-1/2}  &\rho_{-3/2\,\,-3/2}  \\
\end{array}\right)
\ln \left(\begin{array}{cccc}
\rho_{3/2\,\,3/2} & \rho_{3/2\,\,1/2} &\rho_{3/2\,\,-1/2}  &\rho_{3/2\,\,-3/2}  \\
\rho_{1/2\,\,3/2} & \rho_{1/2\,\,1/2} &\rho_{1/2\,\,-1/2} &\rho_{1/2\,\,3/2} \\
\rho_{-1/2\,\,3/2} & \rho_{-1/2\,\,1/2} & \rho_{-1/2\,\,-1/2} & \rho_{-1/2\,\,-3/2} \\
\rho_{-3/2\,\,3/2} & \rho_{-3/2\,\,1/2} &\rho_{-3/2\,\,-1/2}  &\rho_{-3/2\,\,-3/2}  \\
\end{array}\right)\leq\nonumber\\
&&-\mbox{Tr}\left(\begin{array}{cc}
\rho_{3/2\,\,3/2}+ \rho_{1/2\,\,1/2} &\rho_{3/2\,\,-1/2}+ \rho_{1/2\,\,-3/2} \\
\rho_{-1/2\,\,3/2}+\rho_{-3/2\,\,1/2} &\rho_{-1/2\,\,-1/2}+ \rho_{-3/2\,\,-3/2}\\
\end{array}\right)
\ln \left(\begin{array}{cc}
\rho_{3/2\,\,3/2}+ \rho_{1/2\,\,1/2} &\rho_{3/2\,\,-1/2}+ \rho_{1/2\,\,-3/2} \\
\rho_{-1/2\,\,3/2}+\rho_{-3/2\,\,1/2} &\rho_{-1/2\,\,-1/2}+ \rho_{-3/2\,\,-3/2}\\
\end{array}\right)\nonumber\\
 &&-\mbox{Tr}\left(\begin{array}{cc}
\rho_{3/2\,\,3/2}+ \rho_{-1/2\,\,-1/2} &\rho_{3/2\,\,1/2}+ \rho_{-1/2\,\,-3/2} \\
\rho_{1/2\,\,3/2}+\rho_{-3/2\,\,-1/2} &\rho_{1/2\,\,1/2}+ \rho_{-3/2\,\,-3/2}\\
\end{array}\right)
\ln \left(\begin{array}{cc}
\rho_{3/2\,\,3/2}+ \rho_{-1/2\,\,-1/2} &\rho_{3/2\,\,1/2}+ \rho_{-1/2\,\,-3/2} \\
\rho_{1/2\,\,3/2}+\rho_{-3/2\,\,-1/2} &\rho_{1/2\,\,1/2}+ \rho_{-3/2\,\,-3/2}\\
\end{array}\right).\nonumber\\ \label{eq.21a}\end{eqnarray}
The "subadditivity condition" (\ref{eq.21a}) takes place for the system (qudit with $j=3/2$) which has no subsystems.

 Other inequalities of such a form are obtained by arbitrary permutations in (\ref{eq.21a}) of four numbers $3/2,1/2,-1/2,-3/2$. Now we consider system of two particles with $j=1/2$. The spin states are: for spin $j=1\quad$ $|e_1\rangle=|11\rangle,$ $|e_2\rangle=|10\rangle$, $|e_3\rangle=|1-1\rangle,$ and for spin $j=0$ the state is $|e_4\rangle=|00\rangle$. It means that density $4\times4$ - matrix for this system has matrix elements $\rho_{j k}$, where $j$ and $k$ are equal to $1,2,3,4$ and these numbers are mapped onto pairs of numbers $1\Leftrightarrow1\,1$, $2\Leftrightarrow1\,0$, $3\Leftrightarrow1\,-1$, $4\Leftrightarrow0\,0$. In view of this map one can get the entropic subadditivity condition of the form (\ref{eq.18a}) with the introduced substitutions of the indices. In fact, the introduced new inequality for spin $1$ and spin $0$ system of two particles with spins $j=1/2$ can be expressed in terms of the inequality written in the basis $|m m'\rangle$ if one uses the connection of two different basises via Clebsh-Gordon coefficients, providing the unitary transform $4\times4$-matrix
 \begin{equation}\label{eq.22a}
 C=\left(
     \begin{array}{ccc}
       1 & 0 & 0 \\
       0 & F & 0 \\
       0 & 0 & 1 \\
     \end{array}
   \right),
 \end{equation}
where quantum Fourier transform $F$ reads
\begin{equation}\label{eq.23a}
F=\frac{1}{\sqrt2}\left(
                    \begin{array}{cc}
                      1 & 1 \\
                      -1 & 1 \\
                    \end{array}
                  \right).
\end{equation}
The basis $|e_n\rangle$ and basis $|mm'\rangle$ are related as
\begin{equation}\label{eq.24a}
|e_1\rangle=|1/2\,1/2\rangle,\,\,|e_2\rangle=\frac{1}{\sqrt2}(|1/2\,1/2\rangle+|1/2\,-1/2\rangle),\,\,
|e_3\rangle=\frac{1}{\sqrt2}(|1/2\,1/2\rangle-|1/2\,-1/2\rangle),\,|e_4\rangle=|-1/2\,-1/2\rangle.
\end{equation}
Thus, the matrix of density operator $\langle e_n|\hat\rho|e_m\rangle=\rho_{n m}^{(1)}$ and the density matrix $\rho_{m_1m_2m_1'm_2}\equiv\rho_{m_1m_2m_1'm_2'}^{(2)}$ are connected via basis transform expressed in terms of unitary matrix $C$. Both matrices $\rho^{(1)}$ and $\rho^{(2)}$ satisfy the entropic inequality corresponding to subadditivity condition. This property is partial case of general obvious property of two density matrices which are connected by arbitrary unitary transform, i.e. $\rho^{(1)}=u\rho^{(2)}u^+$. Both matrices satisfy the discussed entropic inequalities. 

\section{Improved subadditivety condition}
The tomographic probability distribution for spin states introduced in \cite{DodPLA,OgaJETP,OlgaBregence} and developed in \cite{OlgaJRLR2007,Andreev,JPhys2002,ActaPhysics,OlgaJPhys,Fortsch} provide the possibility to describe the states with density matrix $\rho$ of qudits by means of tomogram
\begin{equation}\label{eq.25a}
w(m,n)=\langle m|u\rho u^+|m\rangle.
\end{equation}
Here $u$ is a unitary matrix, $m=-j,-j+1,\ldots, j,\,\, j=0,1/2,1,\ldots$ are spin projections. The matrix $u$ can be also considered as matrix of irreducible representation of SU(2) - group. For several qudits the tomogram of composite system states reads
\[
w({\bf m},n)=\langle {\bf m}|u\rho(1,2,\ldots,N) u^+|{\bf m}\rangle, \quad {\bf m}=(m_1,m_2,\ldots,m_N).
\]
The vector ${\bf m}$ components $m_k=-j_k,-j_k+1,\ldots,j_k$ correspond to the qudit with spin $j_k$. The tomogram $w({\bf m},u)$ is joint probability distribution of random spin projections $m_1,m_2,\ldots,m_N$ depending on fixed unitary matrix $u$. The matrix can be considered as direct product $u=u_1\otimes u_2\otimes u_N$ of matrices of irreducible representations of the $SU(2)$-group. There is tomographic Shannon entropy \cite{FoundPhysRita,PhysScrRita } corresponding to tomogram $w({\bf m},u)$ which reads
\begin{equation}\label{eq.26a}
H(u)=-\sum_{{\bf m}}w({\bf m}, u)\ln w({\bf m}, u).
\end{equation}
The minimum of the entropy on the unitary group equals to von Neumann entropy \cite{Mendes,MankoMarmo}
\begin{equation}\label{eq.27a}
-\mbox{Tr}\rho(1,2,\ldots,N)\ln\rho(1,2,\ldots,N)=\min_u H(u)=H(u_0).
\end{equation}
For bipartite system $\rho(1,2)$ one has entropic inequality which is subadditivety condition for tomogram $w(m_1,m_2,u)$. It reads
\begin{equation}\label{eq.28a}
H_1(u)+H_2(u)\geq H(u).
\end{equation}
Here
\begin{equation}\label{eq.29a}
H_k(u)=-\sum_{m_k} w_k(m_k,u)\ln w_k(m_k,u), \quad k=1,2
\end{equation}
and
\begin{equation}\label{eq.30a}
w_1(m_1,u)=\sum_{m_2}w(m_1,m_2,u),\quad w_2(m_2,u)=\sum_{m_1}w(m_1,m_2,u).
\end{equation}
The Shannon information  is $I(u)=H_1(u)+H_2(u)-H(u)$. One has two quantum information inequality
\begin{equation}\label{eq.31a}
I_q=S_1+S_2-S(1,2)\geq0
\end{equation}
where $S_k=-\mbox{Tr}\rho(k)\ln\rho(k)$, $k=1,2,$ $\rho_1=\mbox{Tr}_2\rho(1,2)$, $\rho_2=\mbox{Tr}_1\rho(1,2)$ and the relations
\begin{equation}\label{eq.32a}
S_1=\min_{u_1}H_1(u_1)=H_1(u_{10}),\quad S_2=\min_{u_2}H_2(u_{2})=H_2(u_20).
\end{equation}
The Shannon entropies $H_k(u_k)$ satisfy the equality 
\begin{equation}\label{eq.33a}
H_1(u_1)\equiv H_1(u_1\times u_2), \quad H_2(u_2)\equiv H_2(u_1\times u_2).
\end{equation}
The unitary matrices $u_1$ and $u_2$ are unitary local transforms in the Hilbert spaces of the first and second subsystems, respectively. It was shown \cite{MendesJRLR} that
\begin{equation}\label{eq.34a}
S_1+S_2\geq H(u_{10}\times u_{20})\geq S(1,2).
\end{equation}
Here $S_1=H_1(u_{10}),\, S_2=H_2(u_{20})$.
In \cite{LiebSeiringer} there was obtained the general inequality for entropies.
In tomographic picture of qudit states we can obtain the inequality in the form
\begin{eqnarray}
&&I_q=S_1+S_2-S(1,2)\geq-\large(\sum_{m_1}w_1(m_1,u_1)\ln w_1(m_1,u_1)\large)-\large(\sum_{m_2}w_2(m_2,u_2)\ln w_2(m_2,u_2)\large)\nonumber\\
&&+\large(\sum_{m_1m_2}w(m_1,m_2,u_1\times u_2)\ln w(m_1,m_2,u_1\times u_2)\large).\label{eq.35a}
\end{eqnarray}
It means that quantum information bounds the tomographic information for anu local unitary transforms. 
This inequality gives the inequality associated with the tomographic quantum discord property \cite{RitaJRLR,Yurkevich}
\begin{equation}\label{eq.36a}
S_1+S_2-S_{12}\geq S_1+S_2+\sum_{m_1m_2}w(m_1,m_2,u_1\times u_2)\ln w(m_1,m_2,u_1\times u_2).
\end{equation}
It can be written in the form of spin-tomographic entropic inequality for the case where the unitary matrices of local transforms $u_1$ and $u_2$ are matrices of irreducible representations of the group $SU(2)$. In this case $w(m_1,m_2,u_1\times u_2)\equiv w(m_1,m_2,{\bf n_1},{\bf n_2})$ and
\begin{equation}\label{eq.37a}
w_1(m_1,u_1)\equiv w_1(m_1,{\bf n_1}),\quad w_2(m_2, u_2)\equiv w_2(m_2,{\bf n_2}),
\end{equation}
where ${\bf n_1}$ and ${\bf n_2}$ are unit vectors ${\bf n_1}^2={\bf n_2}^2=1$ determining the directions of spin projection axes.

The tomographic entropies $H_k(u_k)$ and $H(u)$ become the functions of the sphere $S^2$, i.e.
\begin{equation}\label{eq.38a}
H_1(u_1)\equiv H_1({\bf n}_1),\quad H_2(u_2)\equiv H_2({\bf n}_2), \quad H(u_1\times u_2)\equiv H({\bf n}_1,{\bf n}_2).
\end{equation}
The inequality (\ref{eq.35a}) reads
\begin{equation}\label{eq.39a}
S_1+S_2-S(1,2)\geq\langle I\rangle,
\end{equation}
where
\begin{equation}\label{eq.41a}
\langle I\rangle=\frac{1}{4\pi}\int H_1({\bf n}_1) d{\bf n}_1+\frac{1}{4\pi}\int H_2({\bf n}_2) d{\bf n}_2-\frac{1}{16\pi^2}\int H({\bf n}_1,{\bf n}_2)\, d{\bf n}_1\,d{\bf n}_2
\end{equation}
is the averaged spin-tomographic information. The difference of quantum information $I_q$ and maximum of the unitary tomographic information $I_t$
\begin{equation}\label{eq.42a}
I_t=\max_{u_1\times u_2}[H_1(u_1)+H_2(u_2)-H(u_1\times u_2)],
\end{equation}
i.e.
\begin{equation}\label{eq.43a}
I_q-I_t=\Delta I\geq0
\end{equation}
is a characteristics of quantum correlations in bipartite qudit system. One has information inequality
\begin{equation}\label{eq.44a}
I_q-\langle I\rangle\geq\Delta I\geq 0.
\end{equation}
\section{Nonlinear positive maps and new inequalities for qudits}
We study now another inequality which follows from the results of the work \cite{LiebSeiringer}. Let us construct nonlinear positive maps of the density $4\times4$ - matrix $\rho$ of the form
\begin{equation}\label{eq.45a}
\rho\rightarrow\rho^{(1)}\rightarrow\frac{1}{\rho_{11}+\rho_{22}}\left(
                                                                   \begin{array}{cc}
                                                                     \rho_{11}& \rho_{12} \\
                                                                     \rho_{21} &\rho_{22} \\
                                                                   \end{array}
                                                                 \right);\quad
 \rho\rightarrow\rho^{(2)}\rightarrow\frac{1}{\rho_{33}+\rho_{44}}\left(
                                                                   \begin{array}{cc}
                                                                     \rho_{33}& \rho_{34} \\
                                                                     \rho_{43} &\rho_{44} \\
                                                                   \end{array}
                                                                 \right)
\end{equation}
as well as the linear positive maps $\rho\rightarrow\rho_1$ and $\rho\rightarrow\rho_2$, where $\rho_1$ and $\rho_2$ are determined by (\ref{eq.14a}). It turns out that there exist new inequality which is valid for arbitrary density $4\times4$-matrix/ The inequality can be derived from the results of \cite{LiebSeiringer}. For density $4\times4$ matrix with elements $\rho_{jk},\,j,k=1,2,3,4$ it reads
\begin{eqnarray}\label{eq.47a}
&&\langle S_2\rangle=-\mbox{Tr}\Large[\left(\begin{array}{cc}\rho_{11}& \rho_{12} \\
                                                                     \rho_{21} &\rho_{22} \\
                                                                   \end{array}
                                                                 \right)
\ln\left(\begin{array}{cc}\frac{\rho_{11}}{\rho_{11}+\rho_{22}}& \frac{\rho_{12}}{\rho_{11}+\rho_{22}} \\
                                                                     \frac{\rho_{21}}{\rho_{11}+\rho_{22}} &\frac{\rho_{22}}{\rho_{11}+\rho_{22}} \\
                                                                   \end{array}
                                                                 \right)\Large]
-\mbox{Tr}\Large[\left(
                                                                   \begin{array}{cc}
                                                                     \rho_{33}& \rho_{34} \\
                                                                     \rho_{43} &\rho_{44} \\
                                                                   \end{array}
                                                                 \right)\ln\left(
                                                                   \begin{array}{cc}
                                                                     \frac{\rho_{33}}{\rho_{33}+\rho_{44}}&\frac{ \rho_{34}}{\rho_{33}+\rho_{44}} \\
                                                                     \frac{\rho_{43}}{\rho_{33}+\rho_{44}} &\frac{\rho_{44}}{\rho_{33}+\rho_{44}} \\
                                                                   \end{array}
                                                                 \right)\Large]\nonumber\\
&&\leq-\mbox{Tr}\Large[\left(
             \begin{array}{cc}
               \rho_{11}+\rho_{33} &\rho_{12}+\rho_{34} \\
               \rho_{21}+\rho_{43} &\rho_{22}+\rho_{44} \\
             \end{array}
           \right)\ln\left(
             \begin{array}{cc}
               \rho_{11}+\rho_{33} &\rho_{12}+\rho_{34} \\
               \rho_{21}+\rho_{43} &\rho_{22}+\rho_{44} \\
             \end{array}
           \right)\Large].\label{eq.47a}
\end{eqnarray}
The inequality (\ref{eq.47a}) cam be applied for density matrices of two-qubit states, for density matrix of qudit with $j=3/2$ and for states of two qubits in the basis of states of spin $j=1$ and $j=0$. For this we apply the corresponding notations of numbers $1,2,3,4$ discussed in previous sections. The inequality (\ref{eq.47a}) means that there are two inequalities of the form \cite{MendesJRLR}
\begin{equation}\label{eq.48a}
S_1+S_2\geq H(u_{10}\times u_{20}\geq S(1,2)
\end{equation}
and the inequality
\begin{equation}\label{eq.49a}
S_1+S_2\geq S_1+\langle S_2\rangle\geq S(1,2).
\end{equation}
The inequality (\ref{eq.47a}) can be extended to get the inequalities by means of permutations of the numbers $1,2,3,4.$ The analogous inequalities can be written for arbitrary composite qudit systems including only one qudit. 

\section{Conclusions}
\pst To conclude we present the main results of our study. We obtained new inequalities for the composite systems of qudits and for system without subsystems. We obtained the condition analogous to subadditivity condition for qudit with spin $j=3/2$. Also we obtained new inequality which is analog of "stronger" subadditivity condition for systems without subsystems. The entropic inequality example of $j=3/2$ states and $j=1\oplus j=0$ states as well as  two-qubit states were studied in a unified method. The method can be used to formulate some generic entropic inequalities for arbitrary Hermitian nonnegative trace-class matrices. It was pointed out that all the discussed inequalities take place for the matrices independently on the product structure of the Hilbert space. This means that all the information entropic inequalities are valid for both the composite and noncomposite quantum systems.


\end{document}